\documentclass[a4paper]{aa}
\usepackage{natbib}
\usepackage{txfonts}
\bibpunct{(}{)}{;}{a}{}{,}
\usepackage{amssymb}

\newcommand{\z}[0]{{\mathbf z}}
\newcommand{\ie}[0]{{i.e.}, }
\newcommand{\eg}[0]{{e.g.}, }
\usepackage{graphics}

\newcommand{\image}[4]
{
\begin{figure}[htb]
\centering
\makebox{
  \resizebox
      {#2cm}{!}
      {\includegraphics{#1}}
}
\caption{#4}
\label{#3}
\end{figure}
}

\begin{document}

\title{Kinematics of Extragalactic Radio Sources}
\subtitle{A Model for Symmetric Jets}

\author{M. Thulasidas\inst{1}}
\offprints{M. Thulasidas}
\institute{New Initiatives, Institute for Infocomm Research,
21 Heng Mui Keng Terrace, Singapore 119613.\\
\email{manoj@i2r.a-star.edu.sg}}
\date{Submitted to A\&A: November 29, 2004; AA/2004/2445}

\abstract { Symmetric extragalactic radio sources exhibit a baffling
  array of features that are very poorly understood.  What is the
  origin of the remarkable symmetry between the two lobes?  Why is the
  emission in the radio frequency (RF) range?  Why are the jets so
  well collimated, implying long term memory of the core?  Why do the
  core regions emit blue and ultraviolet light?  What is the origin of
  transient X--ray and $\gamma$ ray bursts?  In this article, we present a
  model for the kinematics of extragalactic radio sources that
  explains these features. First, we show that the traditional
  explanation \citep{rees} for the apparent superluminal motion (which
  is based on the light travel time effect) requires the sources to be
  highly asymmetrical.  This is in stark contrast with the observation
  of most radio sources with double lobed structures, which tend to
  look very symmetrical. We establish that an apparent superluminal
  motion in {\em both\/} the jets of these radio sources (galactic or
  extragalactic) necessarily implies {\em real\/} superluminal motion.
  The light travel time effect influences the way we perceive
  superluminal motion.  An object, moving across our field of vision
  at superluminal speeds, will appear to us as two objects receding
  from a single point.  Based on this effect, we derive the
  kinematical properties of extragalactic radio sources and explain
  the puzzling features listed above.  Furthermore, we derive the time
  evolution of the hot spots and compare it to the proper motions
  reported in the literature \citep{M87}. We also compare the time
  evolution of a microquasar \citep{superluminal3} with our model and
  show excellent agreement.  This model can also explain the observed
  blue/UV spectrum (and its time evolution) of the core region and the
  RF spectrum of the lobes, and why the radio sources appear to be
  associated with galactic nuclei. We make other quantitative
  predictions, which can be verified.


}
\maketitle

\section{Introduction}

\image{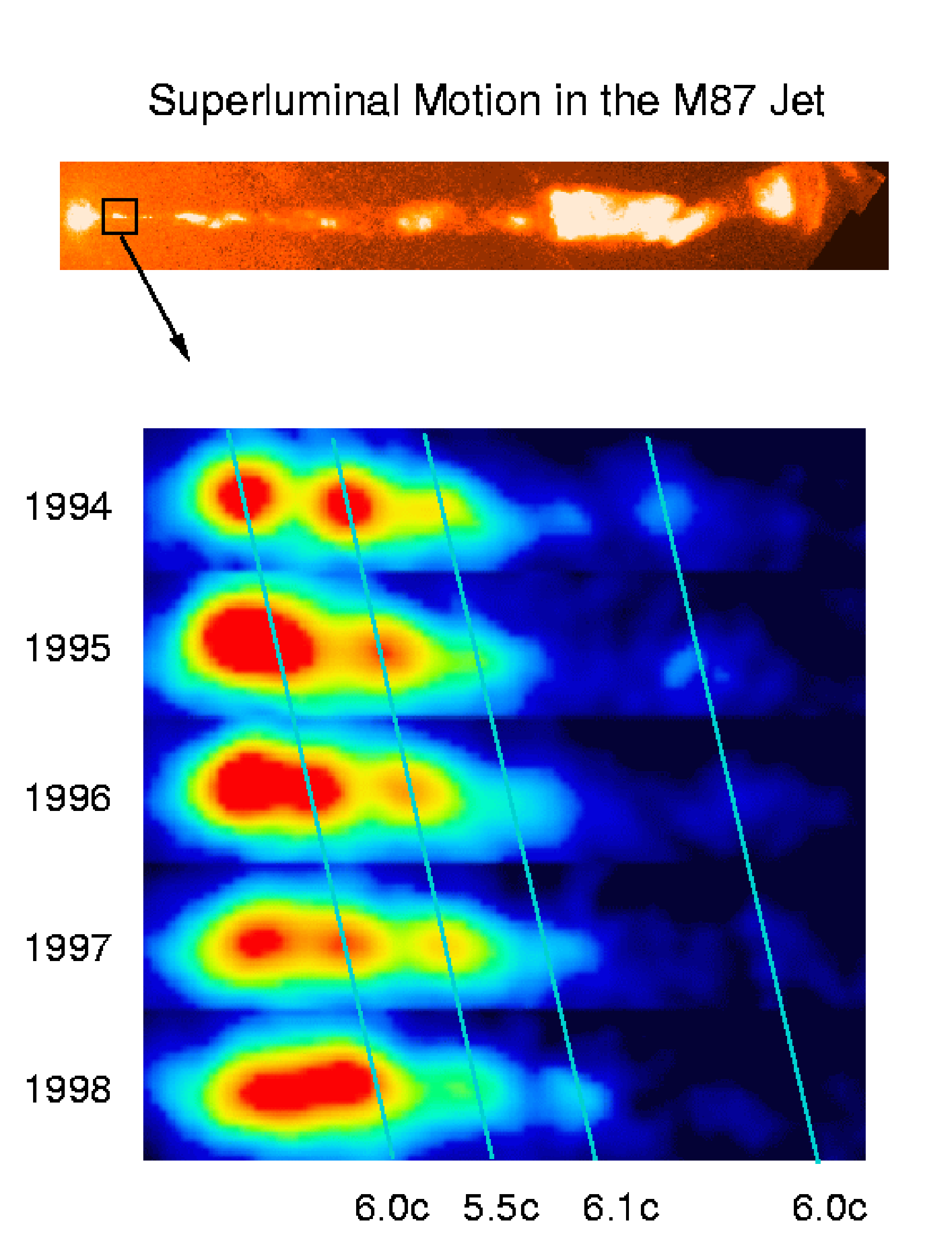}{8}{superluminal} {Sequence of Hubble images showing
  apparent motion at six times the speed of light in the galaxy M87.
  Photo Credit: John Biretta, Space Telescope Science Institute.}

Transverse velocities of a celestial object can be measured almost
directly using angular measurements.  The angular rate can be
translated to a speed using the known (or estimated) distance of the
object from us.  In the past few decades, scientists have observed
\citep{M87,superluminal2} objects moving at apparent transverse
velocities significantly higher than the speed of light.
Fig.~\ref{superluminal} shows one such example.  It shows the image of
the galaxy M87 from the Space Telescope Science Institute.  The top
panel is an image from the Hubble space telescope showing a 5000 light
year long jet from the galaxy's nucleus.  The panel below shows a
sequence of Hubble images showing highly superluminal motion, with the
slanting lines tracking the moving features \citep{M87}. Some such
superluminal objects were detected within our own galaxy
\citep{superluminal1,superluminal3,superluminal4,GRS1915}.

The special theory of relativity \citep{einstein} states that nothing
can accelerate past the speed of light.  A direct measurement of
superluminal objects emanating from a single point (or a small region)
would be a violation of the special theory of relativity at a
fundamental level.  However, before proclaiming a contradiction based
on the measurement of an apparent superluminal motion, one has to
establish that the measurement is not an artifact of the way one
perceives transverse velocities.  \citet{rees} offered an explanation
why such apparent superluminal motion is not in disagreement with the
special theory of relativity, even before the phenomenon was
discovered.  When an object travels at a high speed towards an
observer, at a shallow angle with respect to his line of sight, it can
appear to possess superluminal speeds.  Thus, a measurement of
superluminal transverse velocity by itself is not an evidence against
the special theory of relativity.  In this article, we re-examine this
explanation (also known also as the light travel time effect.)  We
will show that the apparent symmetry of the extragalactic radio
sources is inconsistent with this explanation.

\image{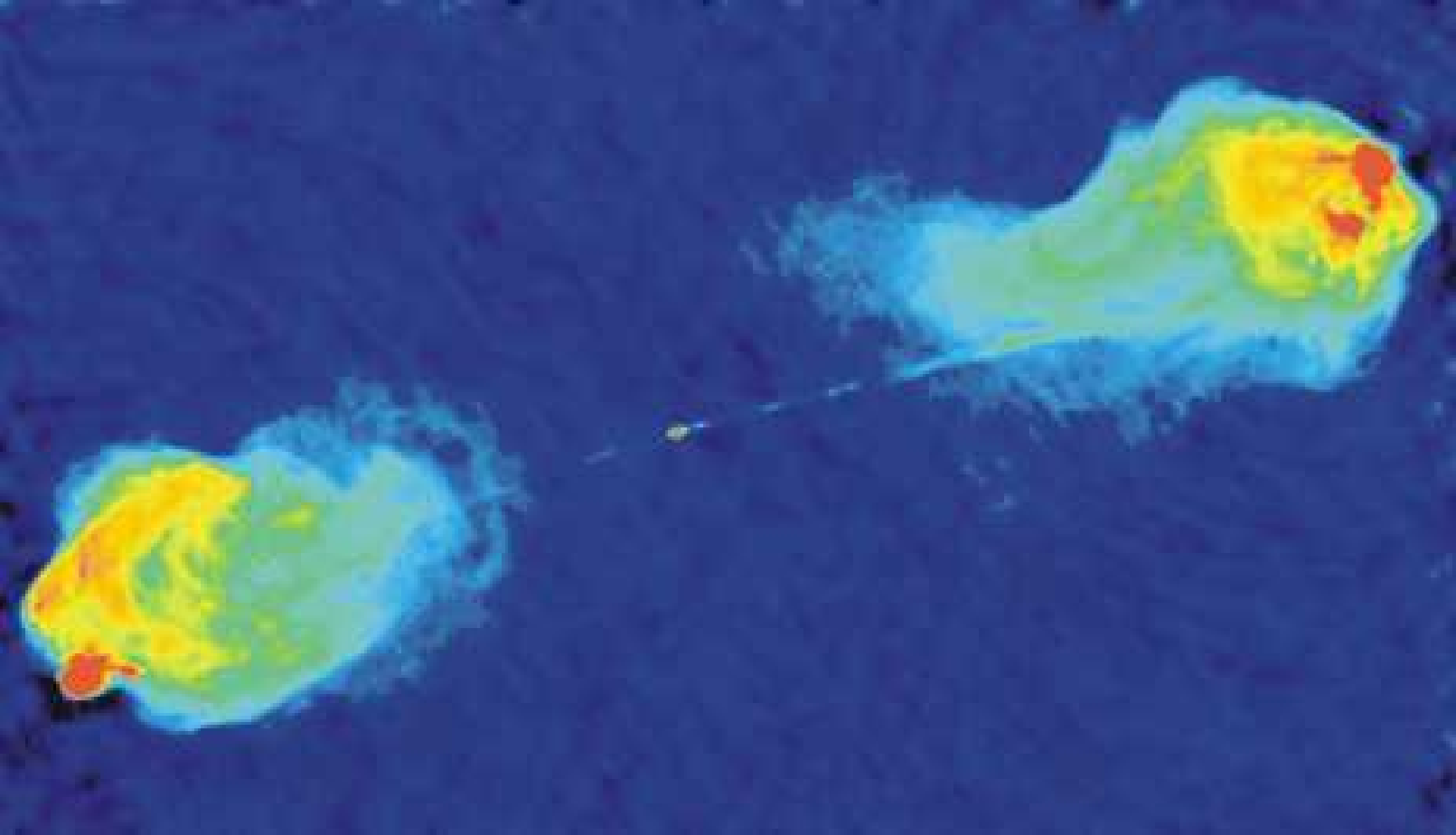}{8}{cyga} {False colour image of the radio jet and
  lobes in the hyperluminous radio galaxy Cygnus A. Red shows regions
  with the brightest radio emission, while blue shows regions of
  fainter emission.  Image courtesy of NRAO/AUI.}

Transverse superluminal motions are usually observed in quasars and
microquasars. Different classes of such objects associated with Active
Galactic Nuclei (AGN) were found in the last fifty years.
Fig.~\ref{cyga} shows the radio galaxy, Cygnus A \citep{cyga} -- one
of the brightest radio objects.  Many of its features are common to
most extragalactic radio sources -- the symmetric double lobes, an
indication of a core, an appearance of jets feeding the lobes and the
hot spots.  \citet{hotspot1} and \citet{hotspot2} have reported more
detailed kinematical features, such as the proper motion of the hot
spots in the lobes.  We will show that our perception of an object
crossing our field of vision at a constant superluminal speed is
remarkably similar to a pair of symmetric hot spots departing from a
fixed point with a decelerating rate of angular separation.  We will
make other quantitative predictions that can be verified, either from
the current data or with dedicated experiments.

\section{Traditional Explanation}

\image{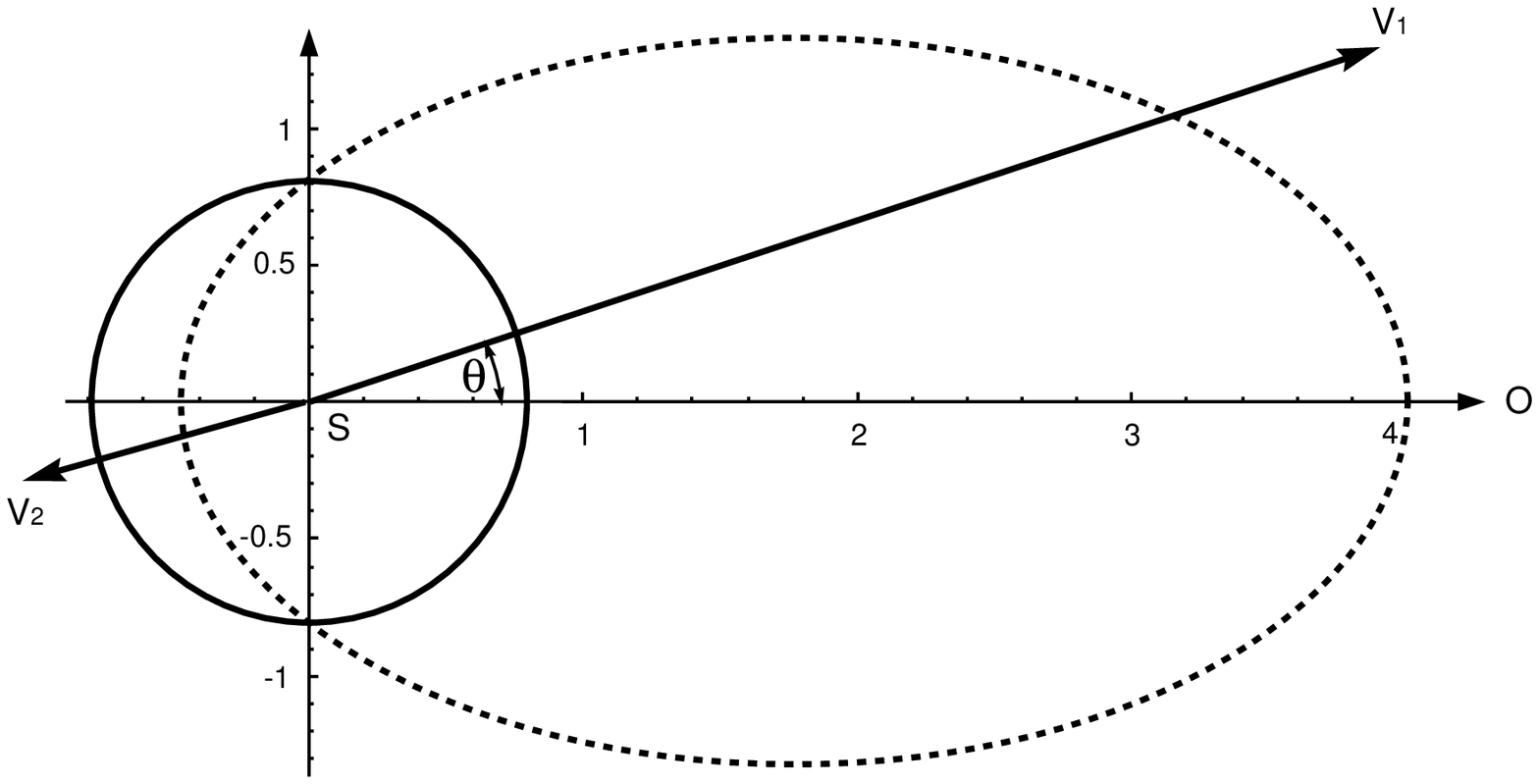}{8}{rees1} {Illustration of the traditional
  explanation for the apparent superluminal motion.  An object
  expanding at a speed $\beta = 0.8$, starting from a single point S.
  The solid circle represents the boundary one second later.  The
  observer is far away on the right hand side, O ($x\to\infty$).  The
  dashed ellipse is the apparent boundary of the object, as seen by
  the observer.}

First, we look at the traditional explanation of the apparent
superluminal motion by the light travel time effect.  Fig.~\ref{rees1}
illustrates the explanation of apparent superluminal motion as
described in the seminal paper by \citet{rees}.  In this figure, the
object at S is expanding radially at a constant speed of $0.8c$, a
highly relativistic speed.  The part of the object expanding along the
direction $V_1$, close to the line of sight of the observer, will
appear to be travelling much faster.  This will result in apparent
transverse velocity that can be superluminal.

The apparent speed $\beta'$ of the object depends on the real speed
$\beta$ and the angle between its direction of motion and the
observer's line of sight, $\theta$.  As shown in the appendix on
Mathematical Details,
\begin{equation}
\beta' \quad=\quad \frac{\beta}{1\,-\,\beta \cos\theta} \label{eqn.1}
\end{equation}
Fig.~\ref{rees1} is a representation of equation~(\ref{eqn.1}) as
$\cos\theta$ is varied over its range.  It is the locus of $\beta'$
for a constant $\beta = 0.8$, plotted against the angle $\,\theta$.
The predicted shape of the apparent speed is in complete agreement
with what was predicted in 1966 (Fig.~1 in that article \citep{rees}).

For a narrow range of $\,\theta$, the transverse component of the
apparent velocity ($\,\beta'\sin\theta\,$) can appear superluminal.
From equation~(\ref{eqn.1}), it is easy to find this range:
\begin{equation}
\frac{1 - \sqrt{2 \beta^2-1}}{2\beta} < \cos\theta < \frac{1 +
  \sqrt{2 \beta^2-1}}{2\beta} \label{eqn.2}
\end{equation}
Thus, for appropriate values of $\,\beta (>\frac{1}{\sqrt{2}})\,$ and
$\,\theta\,$ (as given in equation~(\ref{eqn.2})), the transverse
velocity of an object can seem superluminal, even though the real
speed is in conformity with the special theory of relativity.

While equations~(\ref{eqn.1}) and (\ref{eqn.2}) explain the apparent
transverse superluminal motion the difficulty arises in the
recessional side.  Along directions such as $V_2$ in Fig.~\ref{rees1},
the apparent velocity is always smaller than the real velocity.  The
jets are believed to be emanating from the same AGN in opposite
directions.  Thus, if one jet is in the $\cos\theta$ range required
for the apparent superluminal motion (similar to $V_1$), then the
other jet has to be in a direction similar to $V_2$.  Along this
direction, the apparent speed is necessarily smaller than the real
speed, due to the same light travel time effect that explains the
apparent superluminal motion along $V_1$. Thus, the observed symmetry
of these objects is inconsistent with the explanation based on the
light travel time effect.  Specifically, superluminality can never be
observed in both the jets (which, indeed, has not been reported so
far).  However, there is significant evidence of near symmetric
outflows \citep{asymmetry} from a large number of ojbects similar to
Fig.~\ref{cyga}.

One way out of this difficulty is to consider hypothetical
superluminal speeds for the objects making up the apparent jets.  Note
that allowing superluminal speeds is not in direct contradiction with
the special theory of relativity, which does not treat superluminality
at all.  The original derivation \citep{einstein} of the theory of
co-ordinate transformation is based on the definition of simultaneity
enforcing the constancy of the speed of light.  The synchronisation of
clocks using light rays clearly cannot be done if the two frames are
moving with respect to each other at superluminal speeds.  All the
ensuing equations apply only to subluminal speeds.  It does not
necessarily preclude the possibility of superluminal motion. However,
an object starting from a fixed point and accelerating past the speed
of light is clear violation of the special theory of relativity.

Another consequence of the traditional explanation of the apparent
superluminal speed is that it is invariably associated with a blue
shift. As given in equation~(\ref{eqn.2}), the apparent transverse
superluminal speed, can occur only in a narrow region of $\cos\theta$.
In this region, the longitudinal component of the velocity is always
towards the observer, leading to a blue shift.  The existence of blue
shift associated with all superluminal jets has not been confirmed
experimentally.  Quasars with redshifts have been observed with
associated superluminal jets.  Two examples are: quasars 3C 279
\citep{q3c279} with a redshift $\z = 0.536$ and 3C 216 with $\z =
0.67$ \citep{q3c216}.  However, the Doppler shift of spectral lines
applies only to normal matter, not if the jets are made up of plasma,
as currently believed.  Thus, the current model of jets, made up of
plasma collemated by a magnetic field originating from an accretion
disc, can accommodate the lack of blue shift.

\section{A Model for Symmetric Extragalactic Radio Sources}
\subsection{Symmetric Jets}
Accepting hypothetical superluminal speeds, we can clearly tackle the
second consequence of the traditional explanation (namely, the
necessity of blue shift along with apparent superluminal motion.)
However, it is not clear how we perceive superluminal motion, because
the light travel time effect will influence the way we perceive the
kinematics.  In this section, we will show that a single object moving
superluminally, in a transverse direction across our field of vision, will
look like two objects departing from a single point in a roughly
symmetric fashion.

\image{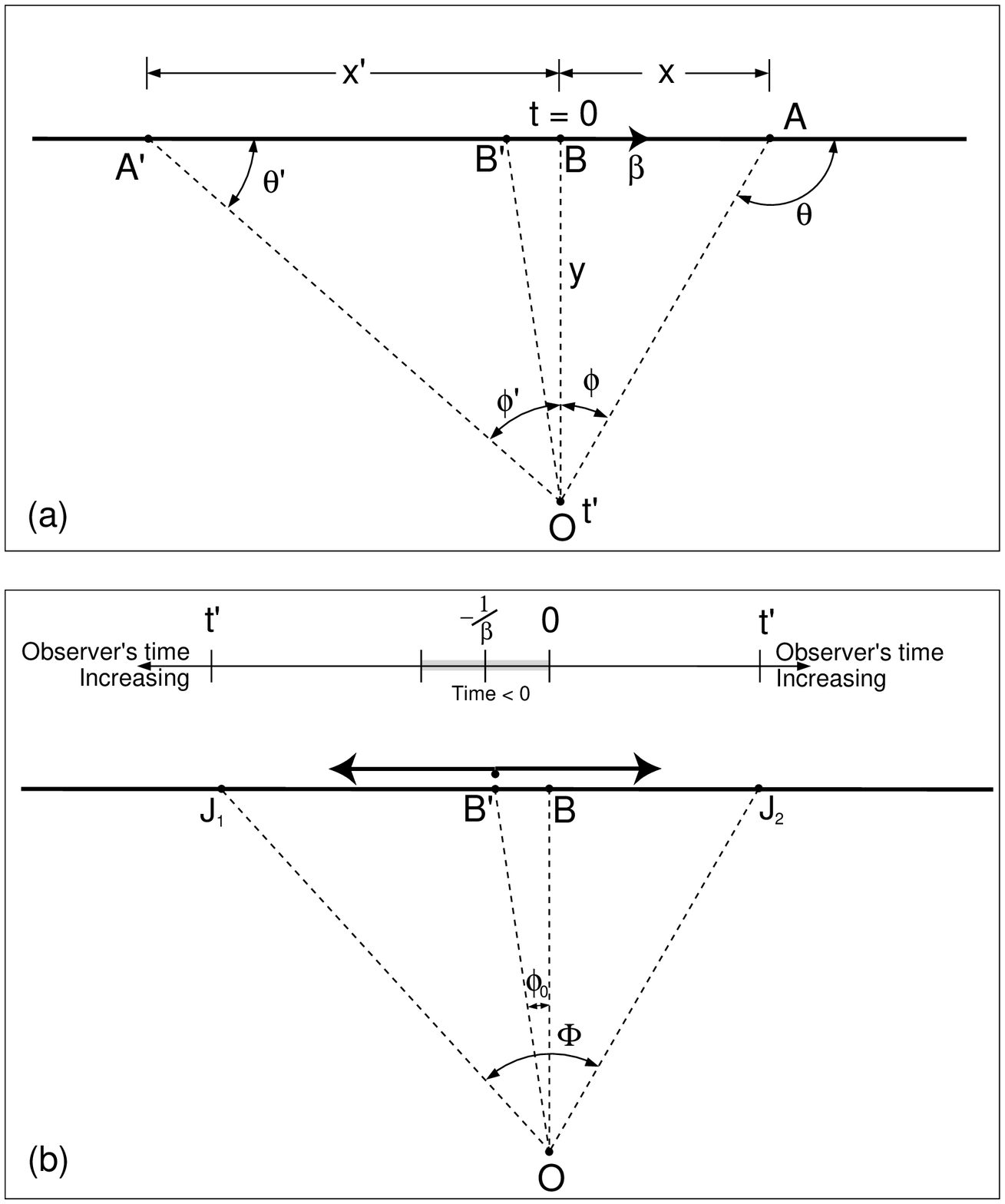}{8}{d} {The top panel (a) shows an object flying along
  $A'BA$ at a constant superluminal speed.  The observer is at $O$.
  The object crosses $B$ (the point of closest approach to $O$) at
  time $t=0$.  The bottom panel (b) shows how the object is perceived
  by the observer at $O$.  It first appears at $B'$, then splits into
  two.  The two apparent objects seem to go away from each other
  (along $J_1$ and $J_2$) as shown.}

Consider an object moving at a superluminal speed as shown in
Fig.~\ref{d}(a).  The point of closest approach is $B$.  At that
point, the object is a distance of $y$ from the observer at $O$. Since
the speed is superluminal, the light emitted by the object at some
point $B'$ (before the point of closest approach $B$) reaches the
observer {\em before\/} the light emitted at $A'$.  This gives an
illusion of the object moving in the direction from $B'$ to $A'$,
while in reality it is moving in the opposite direction.

We use the variable $t'$ to denote the the observer's time.  Note
that, by definition, the origin in the observer's time axis is set
when the object appears at $B$.  $\phi$ is the observed angle with
respect to the point of closest approach $B$. $\phi$ is defined as
$\theta - \pi/2$ where $\theta$ is the angle between the object's
velocity and the observer's line of sight.  $\phi$ is negative for
negative time $t$.

It is easy to derive the relation between $t'$ and $\phi$.  (See
Appendix for the mathematical details.)
\begin{equation}
 t' = y\left( \frac{\tan\phi}\beta + \frac{1}{\cos\phi} - 1\right)
 \label{eqn.3}
\end{equation}
Here, we have chosen units such that $c = 1$, so that $y$ is also the
time light takes to traverse $BO$.  The observer's time is measured
with respect to $y$.  \ie $t' = 0$ when the light from the point of
closest approach $B$ reaches the observer.

The actual plot of $\phi$ as a function of the observer's time is
given in Fig.~\ref{PhiVsTp0}.  Note that for for subluminal speeds,
there is only one angular position for any given $t'$.  The time axis
scales with $y$.  For subluminal objects, the observed angular
position changes almost linearly with the observed time, while for
superluminal objects, the change is parabolic.

Equation~\ref{eqn.3} can be approximated using a Taylor series
expansion as:
\begin{equation}
t' \approx y\left(\frac\phi\beta +
  \frac{\phi^2}{2}\right)\label{eqn.4}
\end{equation}
From the quadratic equation~\ref{eqn.4}, one can easily see that the
minimum value of $t'$ is $t'_{\rm min} = -y/2\beta^2$ and it occurs at
$\phi_{0}=-1/\beta$.  Thus, to the observer, the object first appears
(as though out of nowhere) at the position $\phi_0$ at time $t'_{\rm
  min}$.  Then it appears to stretch and split, rapidly at first, and
slowing down later.  This apparent time evolution of the object is
shown in Fig.~\ref{grs}, where it is compared to the microquasar GRS
1915+105 \citep{superluminal1,GRS1915}.

The angular separation between the objects flying away from each other
is:
\begin{equation}
 \Phi = \frac{2}{\beta}\sqrt{1+\frac{2\beta^2}{y}t'} =
 \frac{2}{\beta}\left(1+\beta\phi\right)
\end{equation}
And the rate at which the separation occurs is:
\begin{equation}
 \frac{d\Phi}{dt'} = \sqrt{\frac{2}{y\Delta t'}} =
 \frac{2\beta}{y\left(1+\beta\phi\right)} \label{eqn.5}
\end{equation}
where $\Delta t' = t' - t'_{\rm min}$, the apparent age of the
symmetric object.

\image{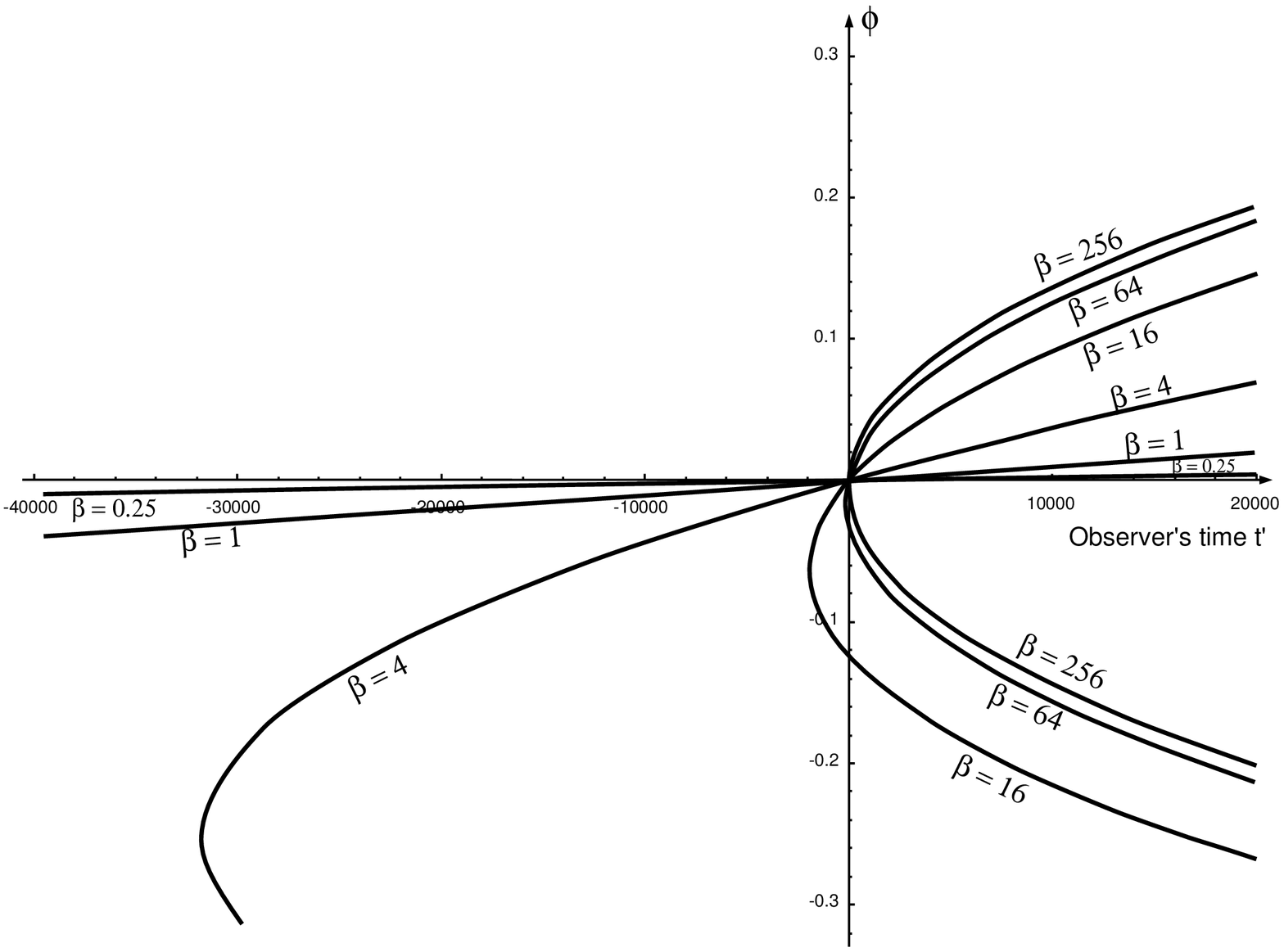}{8}{PhiVsTp0} {The apparent angular positions of an
  object travelling at different speeds at a distance $y$ of one
  million light years from us.  The angular positions ($\phi$ in
  radians) are plotted against the observer's time $t'$ in years.}

\image{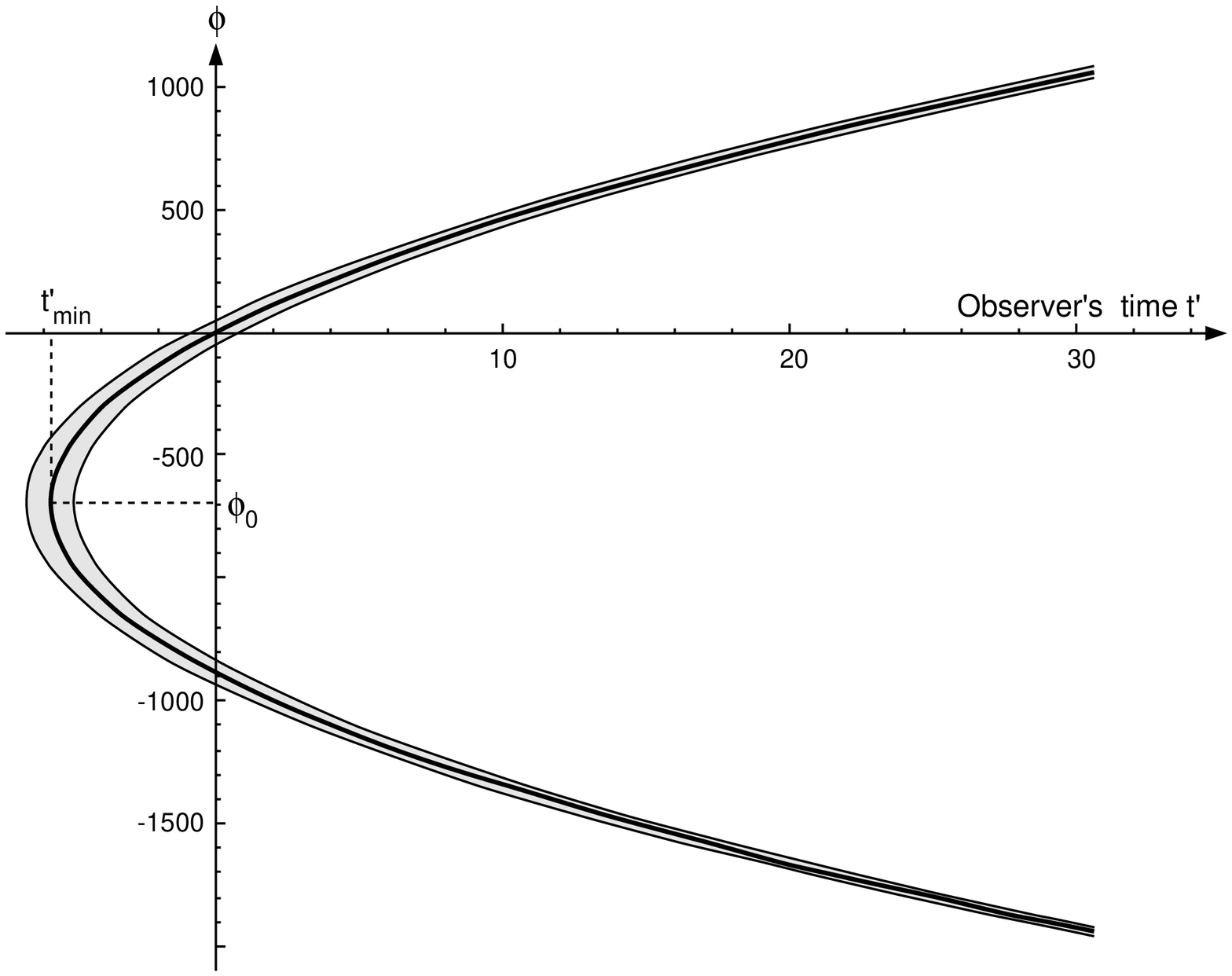}{8}{PhiVsTp} {The apparent angular positions and sizes
  of an object travelling at $\beta = 300$ at a distance $y$ of one
  million light years from us.  The angular positions ($\phi$ in arc
  seconds) are plotted against the observer's time $t'$ in years. The
  shaded region represents the apparent angular spread of the object,
  with an assumed diameter of 500 light years.}

This discussion shows that a single object moving across our field of
vision at superluminal speed creates an illusion of an object
appearing at a at a certain point in time, stretching and splitting
and then moving away from each other.  This time evolution is given in
equation~\ref{eqn.3}, and illustrated in the bottom panel of
Fig.~\ref{d}(b).  Note that the apparent time $t'$ is reversed with
respect to the real time $t$ in the region $A'$ to $B'$.  An event
that happens near $B'$ appears to happen before an event near $A'$.
Thus, the observer may see an apparent violation of causality, but it
is just a part of the light travel time effect.

Fig.~\ref{PhiVsTp} shows the apparent width of a superluminal object
as it evolves.  The width decreases with time, along its direction of
motion.  (See the appendix for the mathematical details.)  Thus, the
appearance is that of two spherical objects appearing out of nowhere,
moving away from each other, and slowly getting compressed into
thinner and thinner ellipsoids and then almost disappearing.

If there are multiple objects, moving as a group, at roughly constant
superluminal speed along the same direction, their appearance would be
a series of objects appearing at the same angular position and moving
away from each other sequentially, one after another.  The apparent
knot in one of the jets always has a corresponding knot in the other
jet.

The calculation presented in this article is done in two dimensions.
If we generalise to three dimensions, we can explain the precession
observed in some systems.  Imagine a cluster of objects, roughly in a
planar configuration (like a spiral galaxy, for instance) moving
together at superluminal speeds with respect to us.  All these objects
will have the points of closest approach to us in small angular region
in our field of vision -- this region is around the point of minimum
distance between the plane and our position.  If the cluster is
rotating (at a slow rate compared to the superluminal linear motion),
then the appearance to us would be the apparent jets changing
directions as a function of time.  The exact nature of the apparent
precession depends on the spatial configuration of the cluster and its
angular speed.

If we can measure the angle $\phi_0$ between the apparent core and the
point of closest approach, we can directly estimate the real speed of
the object $\beta$.  We can clearly see the angular position of the
core.  However, the point of closest approach is not so obvious.  We
will show in the next section that the point of closest approach
corresponds to zero red shift.  (This is obvious intuitively, because
at the the point of closest approach, the longitudinal component of
the velocity is zero.)  If this point ($\phi_0$) can be estimated
accurately, then we can measure the speed directly, from the relation
$\phi_0 = -1/\beta$.

\subsection{Red shifts of the Hot Spots}
We can also work out time evolution of the red shift of the hot spots.
However, as relativistic Doppler shift equation is not defined for
superluminal speeds, we need to work out the relationship between the
redshift ($\z$) and the speed ($\beta$) from first principles.  This
is easily done (see the appendix for the mathematical details):
\begin{equation}
1 \,+\, \z \,=\, \left|1 \,-\, \beta\cos\theta\right| \,=\, \left|1
  \,+\, \beta\sin\phi\right|
\end{equation}
There are two solutions for $\z$, corresponding to the apparent
objects at the two different positions.  However, as shown in the
appendix, they are nearly identical for any given value of the
observer's time, $t'$.

Since we allow superluminal speeds in our model of extragalactic radio
sources, we can explain the radio frequency spectra of the hot spots as
extremely red-shifted blackbody radiation.  The $\beta$s involved in
this explanation are typically very large, and we can approximate the
red shift as:
\begin{equation}
1 \,+\, \z \,\approx\,  \left|\beta\phi\right| \,\approx\,
\frac{\left|\beta\Phi\right|}{2}
\end{equation}
Assuming the object to be a black body similar to the sun, we can
predict the peak wavelength (defined as the wavelength at which the
luminosity is a maximum) of the hot spots as:
\begin{equation}
\lambda_{\rm max} \approx (1+\z) 480nm \approx
\frac{\left|\beta\Phi\right|}{2} 480nm \label{eqn.z}
\end{equation}
where $\Phi$ is the angular separation between the two hot spots.

\image{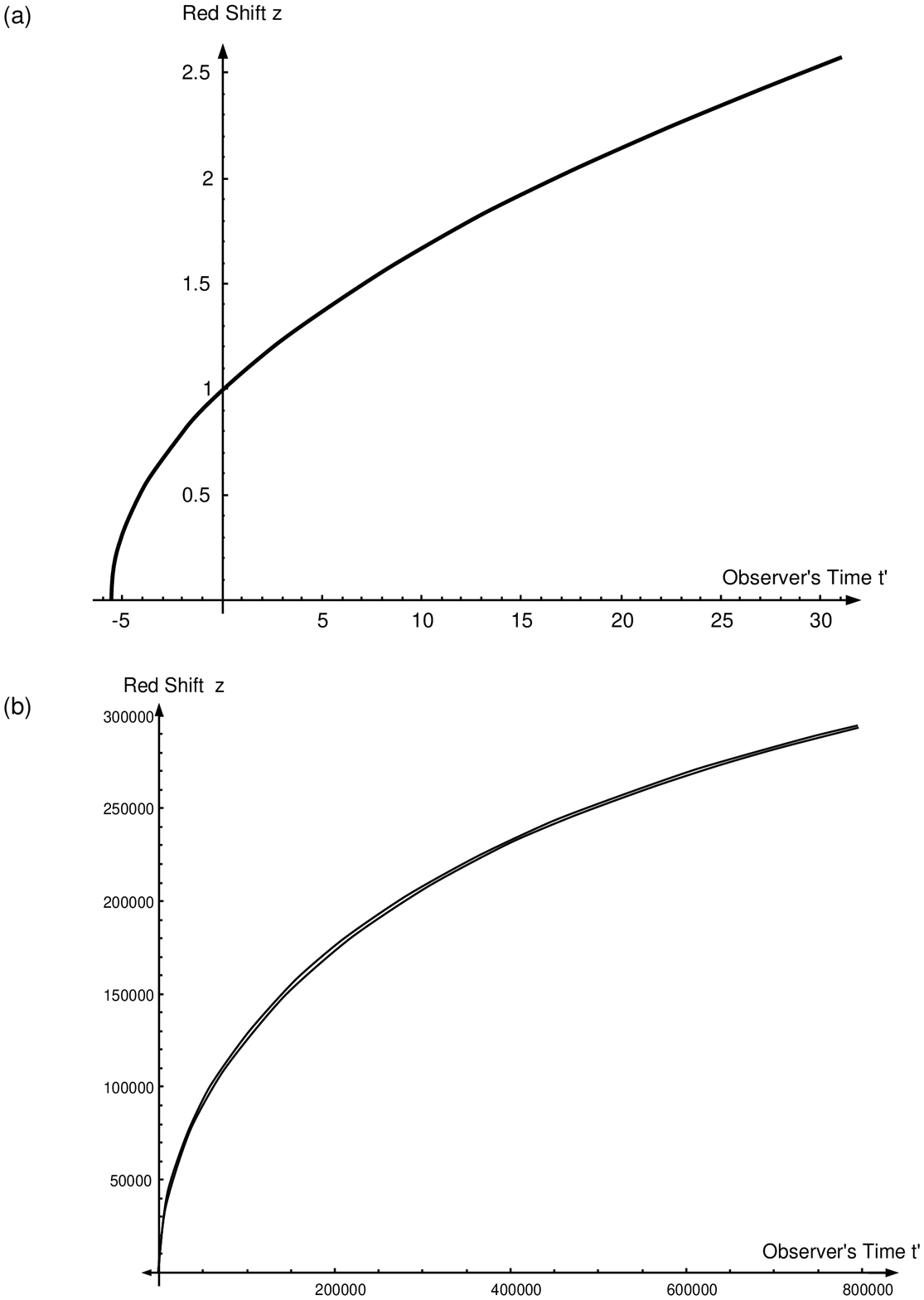}{8}{zVsTp}{Time evolution of the red shift from a
  superluminal object.  The top panel (a) shows the red shifts
  expected from an object moving at $\beta = 300$ at a distance of one
  million light years from us.  In the bottom panel (b), the object is
  moving at a speed $\beta = 300\,000$ at a distance of 15 billion light
  years.  Note that the two apparent objects have nearly identical red
  shifts.}

This shows that the peak RF wavelength increases linearly with the
angular separation.  If multiple hot spots can be located in a twin
jet system, their peak wavelengths will depend only on their angular
separation, in a linear fashion.  The real speed of the single object
masquerading as two hot spots can be estimated from peak wavelength
measurements.  Furthermore, if the measurement is done at a single
radio frequency, intensity variation can be expected as the hot spot
moves along the jet.

Fig.~\ref{zVsTp} shows the variation of red shift as a function of the
observer's time.  In the top panel, we have plotted an object with
$\beta = 300$ and $y =$ one million light years.  For $t' < 0$, there
is a strong blue shift, which explains the observed, transient hard
X--ray spectra of some of the symmetric jets.  \citet{xray1} have a
recent survey of such data, though the currently favoured explanation
for such transient emissions is a stellar tidal disruption scenario.
The small difference between the red shifts of the two apparent
objects may explain the double peak structure observed in the spectral
data of some of the AGNs \citep{doublepeak}.

In order to have a red shift that will push a black body radiation of
a star similar to our sun into RF regions would require a $\beta$ of
about 300\,000 and a distance of closest approach of about 15 billion
light years.  This is plotted in the bottom panel of Fig.~\ref{zVsTp}.

\subsection{Summary of Predictions}
Some of the different quantitative predictions of the model are
recapitulated here.  These are predictions that are relatively easy to
verify with existing data.
\begin{itemize}
\item The appearance of a single object moving across our field of
  vision at superluminal speed is that of an object appearing at a
  point, splitting and moving away in opposite directions.
\item The core will always have a fixed angular position.
\item The new superluminal knots appearing in the jets will always
  appear in pairs.
\item The two apparent objects will shrink monotonically. As the knots
  move away from the core, they become thinner and thinner ellipsoids,
  contracting along the direction of motion while the trasverse size
  remains roughly constant.
\item The separation speed is very high in the beginning, but it slows
  down parabolically with time.
\item The hot spots have almost identical RF spectra (and red shifts).
\item The RF wavelength at which the luminosity of the hot spots is the
  maximum increases linearly with the angular separation between them.
\item Close to the core, the the spectrum is heavily blue shifted.
  Thus, the object can be a strong X--ray or even $\gamma$ ray source
  for a brief period of time.  After that, the spectrum moves through
  optical to RF region.
\end{itemize}

A clear indication of a movement in the core's angular position, or a
superluminal knot appearing without a counterpart in the opposite jet
will be strong evidence against our model based on superluminality.
On the other hand, a clear measurement of apparent superluminal motion
in both the jets (not reported so far) will provide a convincing
indication that the conventional explanation is inadequate.

\subsection{Comparison to Measurements}
\image{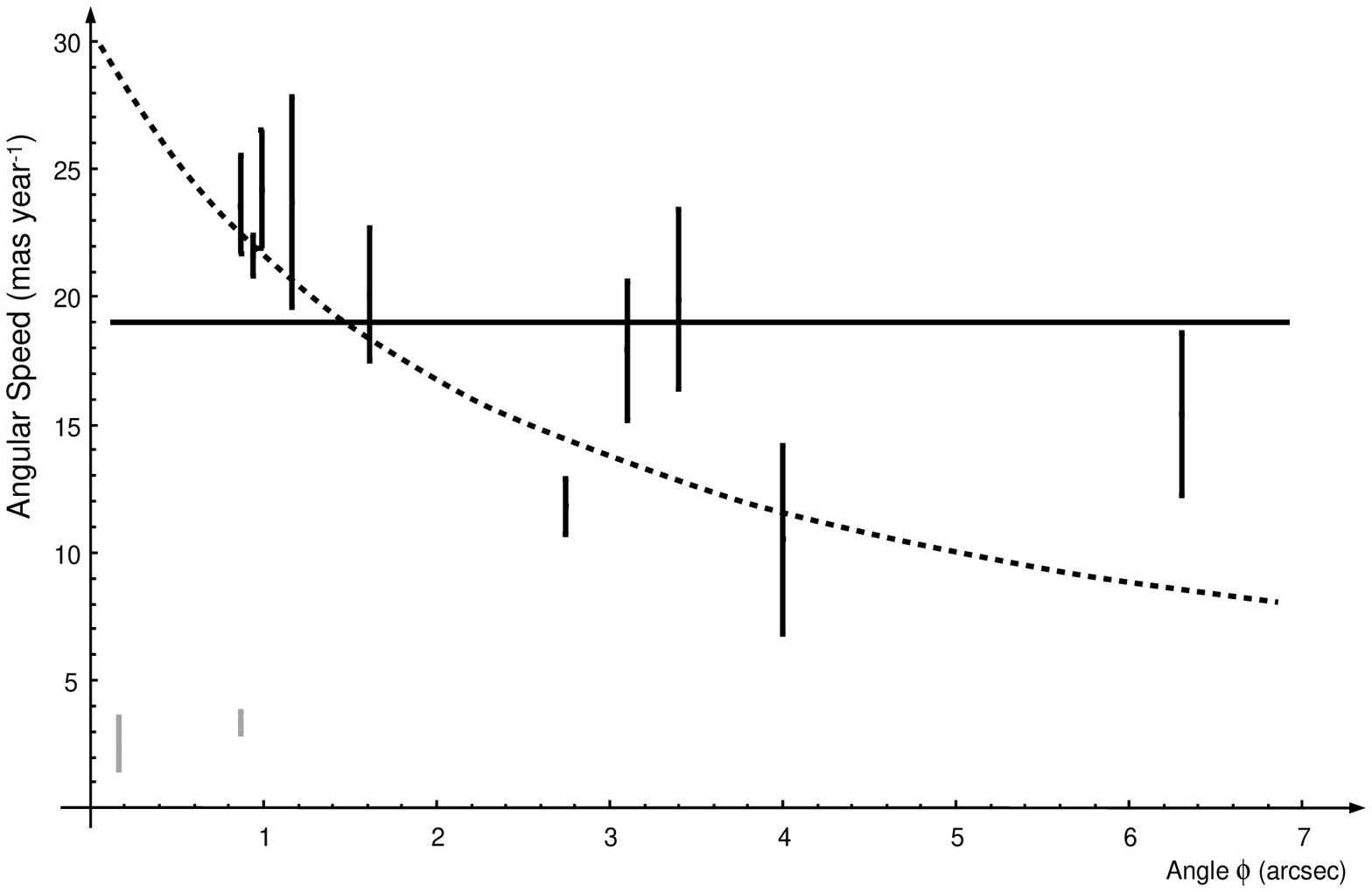}{8}{dphi} {The angular speed of M87 (as a function of
  the feature angle $\phi$) fitted against our model. The solid line
  constrains the distance of closest approach to the estimated
  distance of M87 -- 52 million light years, giving a $\beta = 4.8$.
  The dotted line is a free fit giving $\beta = 84\,000$.}

\citet{M87} have reported proper motion in one of the jets of M87 as a
function of the angle ($\phi$) between the apparent core and the
feature points.  M87 is estimated to be at about 52 million light
years away from us, which gives us the value of $y$. In
equation~\ref{eqn.5}, we have the apparent angular speed ($d\Phi/dt'$)
as a function of $\phi$.  Making the reasonable approximation $\Phi
\approx 2\phi$, we can fit these data to our equation.  The result is
shown in Fig.~\ref{dphi}.  The fit gives a value $\beta = 4.8$.  A
cluster of objects flying across our field of vision, about five times
faster than light and at a distance of 52 million light years, will
look like two jets moving away from each other at roughly 38 mas/year.
If one of the two jets is hidden for some reason, the appearance will
be a single jet of objects moving away from a point with an angular
speed of about 19 mas/year.  Note that we exclude the first two points
from the fit.  In this region close to the core, the appearance of new
objects makes it difficult to track the features.

\image{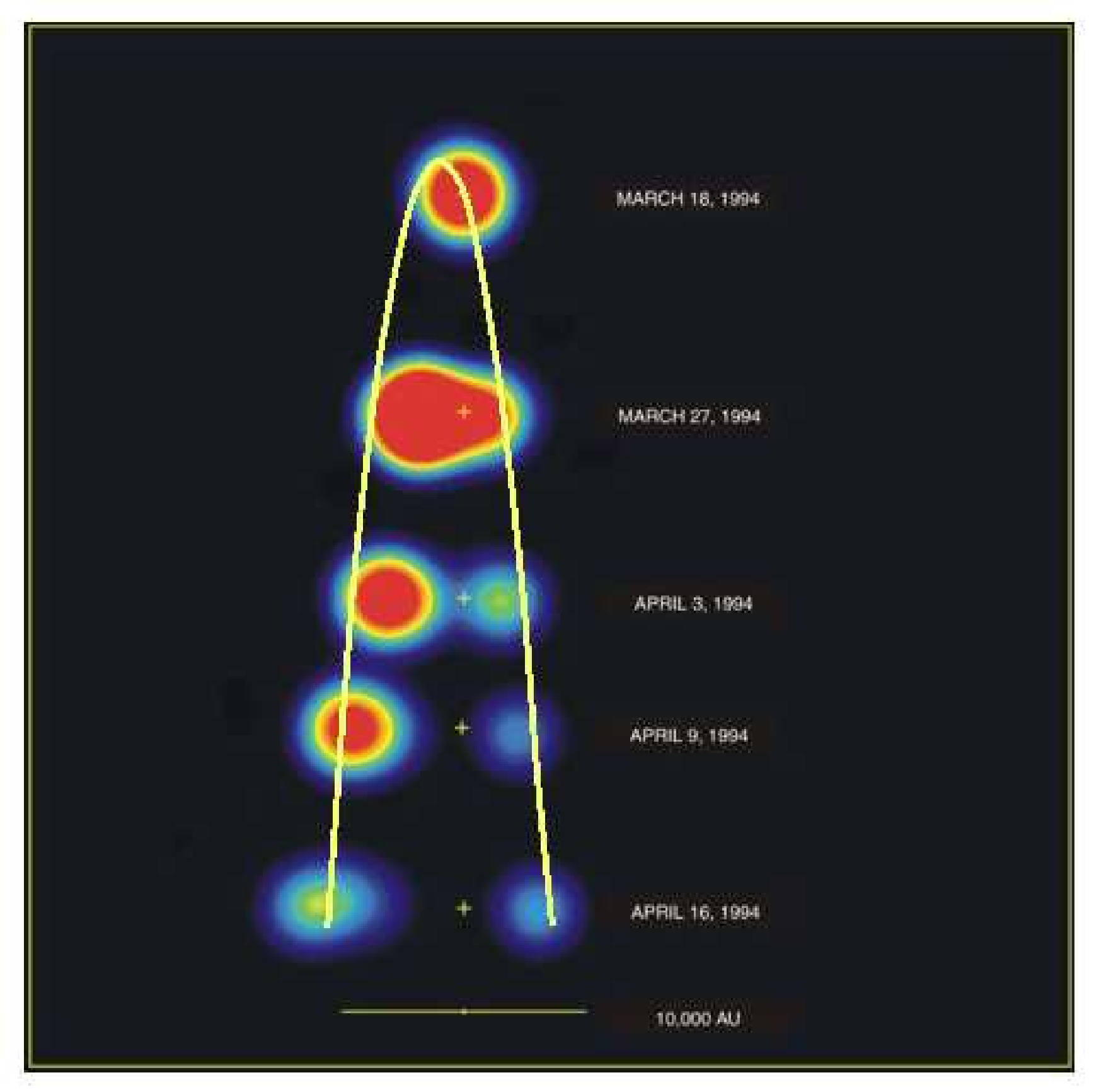}{8}{grs} {Fit of our model to the time evolution
  measurements of GRS1915+105. The yellow curve overlaid corresponds
  to our perception of a single superluminal object, travelling at
  $\beta = 3\,000$ across our field of vision.}

A much better fit can be obtained if we were to let the distance $y$
also float.  The resulting $\beta$ of about 90\,000 may explain the
spectra of the hot spots. While the estimated $\beta$ may look
excessive, once superluminal motion is allowed, there is no a priori
reason why it should not take any value at all.  Fig.~\ref{grs} shows
another comparison of our model to the data available in the
literature.  Here, the time evolution of the microquasar GRS 1915+105
\citep{superluminal1,GRS1915} is fitted to our model of a single
superluminal object. The deceleration of relativistic jets (one of the
predictions from our model) has been observed in the Microquasar XTE
J1550-564 \citep{sci_microquasar}, though it is currently believed to
be an effect similar to frictional drag.

AGNs are known to have intensely blue or ultraviolet core, not easily
explained by thermal models.  But this is an expected feature in our
model. As seen in equation~\ref{eqn.z}, the core (where $\Phi\to0$)
must have a highly blue shifted spectrum. A clear evolution of
emission frequency from ultraviolet to RF is seen in the photometry of
the jet in 3C 273 \citep{RF2UV}.  The spectrum shifts from lower
wavelength to higher as a function of the angular distance from the
core, strikingly consistent with our prediction.

This shifting of peak frequency can be seen at a much larger scale in
Fig.~\ref{cyga}. The size of the optical core is about a tenth of the
angular separation between the hotspots.  If we model Cygnus A as a
collection of objects moving together at superluminal speeds, the core
region would have emissions in the $\gamma$, X-ray, UV or optical
region.  As we move away towards the hotspots, the peak frequency
would continously shift to RF.  This behaviour is indeed reported
\citep{cyga1} recently.  This also partially explains why
extragalectic radio sources seem to be associated with galactic
neuclei, instead of appearing randomly in the sky.  A large collection
of objects moving together (a large spiral galaxy, viewed from the
side, for instance) superluminally gives the impression of a smaller
stationary object with optical emission.  The apparent object is
likely to appear elongated along the direction of motion (with the
major axis along the direction of the jets), with RF lobes appearing
symmetrically farther away from the core.  If the motion is not along
a linear trajectory, we may see curved jets.

\section{Conclusions}
In this article, we explored the full implications of the traditional
explanation for the apparent superluminal motion observed in certain
quasars and microquasars.  The equation that explains the apparent
superluminal speeds predicts that objects receding from us should
appear to be moving slower.  Thus, in a symmetric radio sources where
it is observed, the superluminal motion can appear only in one of the
jets.  The observed symmetry of these extragalactic radio sources
(even subluminal ones) is incompatible with the explanation.  Another
consequence is that an apparent superluminal motion (if the moving
objects are composed of normal matter rather than plasma) must always
show a blue shift, a red shifted object can never be superluminal.

We further explored the possiblity of real superluminal motion.  We
showed that a single superluminal object flying across our field of
vision would appear to us as a symmetric separation of two objects
from a fixed core.  Using this fact as the model for such symmetric
jets, we can explain their kinematical features quantitatively.  In
particular, we showed that the angle of separation of the hot spots is
parabolic in time, and the red shifts of the two hot spots are almost
identical to each other.  Even the fact that the spectra of the hot
spots are in the radio frequency region can be explained by assuming
hyperluminal motion and the consequent red shift of the blackbody
radiation.  Furthermore, the requirement that an apparent superluminal
motion be associated with a blue shift does not apply any more.

We presented a set of predictions and compared them to existing data.
The features such as the blueness of the core, symmetry of the lobes,
the transient $\gamma$ and X-Ray bursts, the measured evolution of the
spectra along the jet all find natural and simple explanations in this
model.  Note, however, that we have not addressed the dynamics of the
model -- how are the super or hyperluminal objects powered?  The only
observation in this article is that a collection of objects travelling
superluminally across our field of vision can appear remarkably
similar to an AGN with symmetrically placed radio frequency lobes and
hotspots.  It does not preclude plasma jets that may be related to
space-time singularities or other massive objects and the associated
accretion discs.  The conventional explanation of the apparent
superluminal motion in asymmetric jets (\eg quasar 3C 279
\citet{q3c279}) also stands.  In fact, our model is just a
generalisation of the conventional explanation.

We argued that superluminal motion is not inconsistent with the
special theory of relativity, which just does not deal with it.
Acceptance of superluminality has far-reaching consequences in other
long established notions of our universe.  In particular, it can be
shown (see Appendix B) that the apparent expansion of the universe at
strictly subluminal speed is also an artifact of our perception of
superluminal motion. Thus, the theory of the big bang will have to be
looked at once again to see how light travel time effect modifies it.
The description of extragalactic (or galactic) radio sources in terms
of superluminal motion has a direct impact on our understanding of
black holes.

\appendix

\section{Mathematical Details}
\image{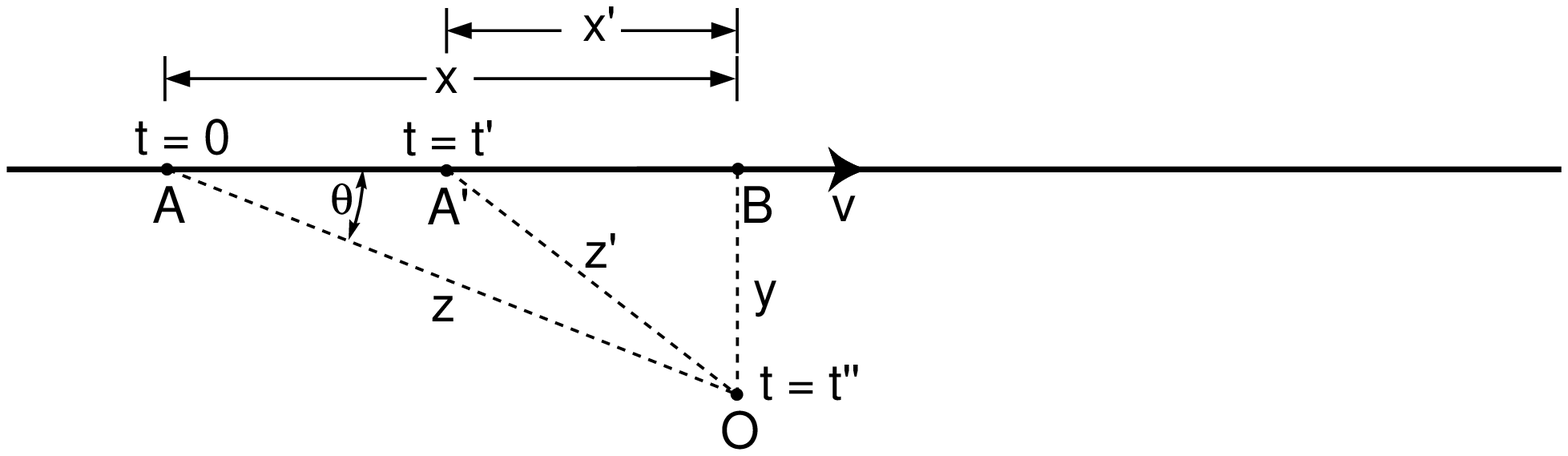}{8}{c} {The object is flying along $AB$, the observer
  is at $O$.  The object crosses $A$ at time $t=0$.  It reaches $A'$
  at time $t = t'$.  A photon emitted at $A$ reaches $O$ at time $t =
  t_0$, and a photon emitted at $A'$ reaches $O$ at time $t = t''$.}

\subsection{Velocity Profile of an Expanding Object}

This section re-derives the ellipse in Fig.~\ref{rees1} from first
principles.  In Fig.~\ref{c}, there is an observer at $O$. An object
is flying by at a high speed $v = \beta c$ along the horizontal line
$AB$.  With no loss of generality, we can assume that $t = 0$ when the
object is at $A$. It passes $A'$ at time $t'$. The photon emitted at
time $t = 0$ reaches the observer at time $t = t_0$, and the photon
emitted at $A'$ (at time $t = t'$) reaches him at time $t = t''$.  The
angle between the object's velocity at $A$ and the observer's line of
sight is $\theta$.  We have the Pythagoras equations:
\begin{eqnarray}
z^2 \,=\, x^2 \,+\, y^2 \quad&\&&\quad z'^2 \,=\, x'^2 \,+\, y^2\\
\Rightarrow\qquad \frac{x+x'}{z+z'} & = & \frac{z-z'}{x-x'} \label{eqn.6}
\end{eqnarray}
If we assume that $x$ and $z$ (distances at time $t_0$) are not very
different from $x'$ and $z'$ respectively (distances at time $t'$), we
can write,
\begin{equation}
\cos\theta = \frac{x}{z} \approx \frac{x+x'}{z+z'} =
\frac{z-z'}{x-x'}
\end{equation}
We define the real speed of the object as:
\begin{equation}
v \,=\, \beta\,c \,=\, \frac{x\,-\,x'}{t'}
\end{equation}
But the speed it {\em appears\/} to have will depend on when the
observer senses the object at $A$ and $A'$.  The apparent speed of the
object is:
\begin{equation}
v' \,=\, \beta'\,c \,=\, \frac{x \,-\, x'}{t'' \,-\,
t_0}
\end{equation}
Thus,
\begin{eqnarray}
\frac{\beta}{\beta'} &=& \frac{t''-t_0}{t'} \\
&=& 1 + \frac{z'-z}{ct'} \\
&=& 1 - \frac{x-x'}{ct'}\cos\theta \\
&=& 1 - \beta\cos\theta
\end{eqnarray}
which gives,
\begin{equation}
\beta' \quad=\quad
\frac{\beta}{1\,-\,\beta \cos\theta}\label{eqn.7}
\end{equation}

Fig.~\ref{rees1} is the locus of $\beta'$ for a constant $\beta =
0.8$, plotted against the angle $\,\theta$.

\subsection{Superluminal Red shift}

Red shift ($\z$) defined as:
\begin{equation}
1 \,+\, \z \,=\, \frac{\lambda'}{\lambda}
\end{equation}
where $\lambda'$ is the measured wavelength and $\lambda$ is the known
wavelength.  In Fig.~\ref{c}, the number of wave cycles created in
time $t'$ between $A$ and $A'$ is the same as the number of wave
cycles sensed at $O$ between $t_0$ and $t''$.  Substituting the
values, we get:
\begin{equation} \frac{t'\, c}{\lambda} \,=\, {\frac{(t''\,-\,t_0)\,c}
{\lambda'}}
\end{equation}
Using the definitions of the real and apparent speeds, it is easy to get
\begin{equation}
\frac{\lambda'}{\lambda} \,=\, \frac{\beta}{\beta'}
\end{equation}
Using the relationship between the real speed $\beta$ and the
apparent speed $\beta'$ (equation~(\ref{eqn.7})),
we get
\begin{equation}
1 \,+\, \z \,=\, \frac{1}{1 \,+\, \beta'\cos\theta} \,=\, 1 \,-\,
\beta\cos\theta
\end{equation}
As expected, $\z$ depends on the longitudinal component of the
velocity of the object.  Since we allow superluminal speeds in this
calculation, we need to generalise this equation for $\z$ noting that
the ratio of wavelengths is positive.  Taking this into account, we
get:
\begin{equation}1 \,+\, \z \,=\, \left|\frac{1}{1 \,+\,
      \beta'\cos\theta}\right|
\,=\, \left|1 \,-\, \beta\cos\theta\right| \label{eqn.8}
\end{equation}

\subsection{Kinematics of Superluminal Objects}
\image{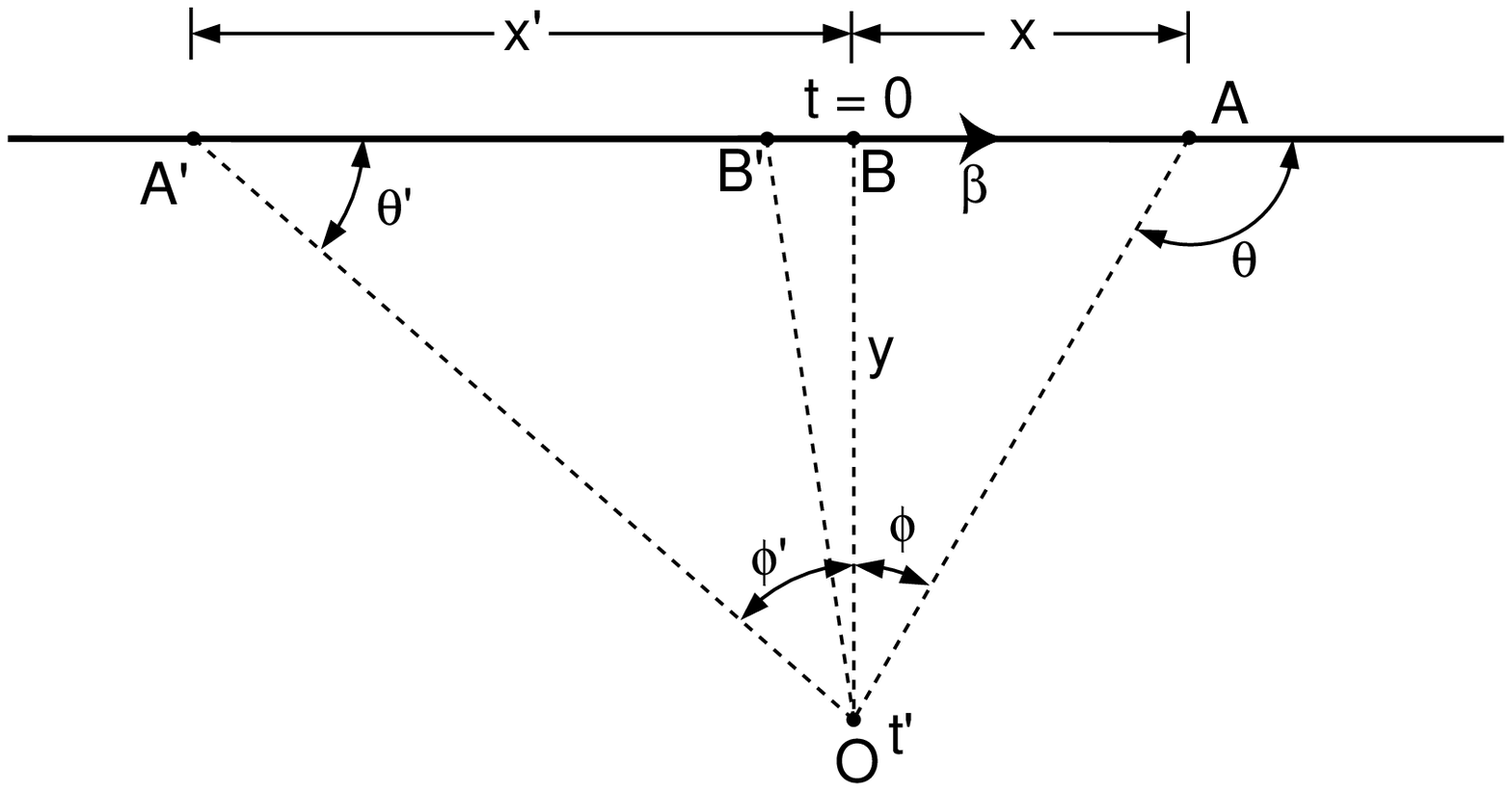}{8}{dd} {An object flying along $A'BA$ at a constant
  superluminal speed.  The observer is at $O$.  The object crosses $B$
  (the point of closest approach to $O$) at time $t=0$.}

The derivation of the kinematics is based on Fig.~\ref{dd}.  Here, an
object is moving at a superluminal speed along $A'BA$. At the point of
closest approach, $B$, the object is a distance of $y$ from the
observer at $O$. Since the speed is superluminal, the light emitted by
the object at some point $B'$ (before the point of closest approach
$B$) reaches the observer {\em before\/} the light emitted at $A'$.
This gives an illusion of the object moving in the direction from $B'$
to $A'$, while in reality it is moving from $A'$ to $B'$.  $\phi$ is
the observed angle with respect to the point of closest approach $B$.
$\phi$ is defined as $\theta - \pi/2$ where $\theta$ is the angle
between the object's velocity and the observer's line of sight.
$\phi$ is negative for negative time $t$.

We choose units such that $c = 1$, in order to make algebra simpler.
$t'$ denotes the the observer's time.  Note that, by definition, the
origin in the observer's time, $t'$ is set when the object appears at
$B$.

The real position of the object at any time $t$ is:
\begin{equation}
 x = y\tan\phi = \beta t
\end{equation}
A photon emitted at $t$ will reach $O$ after traversing the hypotenuse.
A photon emitted at $B$ will reach the observer at $t = y$, since we
have chosen $c = 1$.  If we define the observer's time $t'$ such that
the time of arrival is $t = t' + y$, then we have:
\begin{equation}
 t' = t + \frac{y}{\cos\phi} - y
\end{equation}
which gives the relation between $t'$ and $\phi$.
\begin{equation}
 t' = y\left( \frac{\tan\phi}\beta + \frac{1}{\cos\phi} - 1\right)
\end{equation}
Expanding the equation for $t'$ to second order, we get:
\begin{equation}
 t' = y\left(\frac\phi\beta + \frac{\phi^2}{2}\right)\label{eqn.9}
\end{equation}
The minimum value of $t'$ occurs at $\phi_{0}=-1/\beta$ and it is
$t'_{\rm min} = -y/2\beta^2$.  To the observer, the object first
appears at the position $\phi=-1/\beta$.  Then it appears to stretch
and split, rapidly at first, and slowing down later. The angular
separation between the objects flying away from each other is the
difference between the roots of the quadratic equation~(\ref{eqn.9}):
\begin{equation}
 \Phi = \phi_1-\phi_2 = \frac{2}{\beta}\sqrt{1+\frac{2\beta^2}{y}t'} =
 \frac{2}{\beta}\left(1+\beta\phi\right)
\end{equation}
We also have the mean of the roots equal to the position of the
minimum:
\begin{equation}
\phi_1 + \phi_2 = \frac{-2}{\beta}
\end{equation}
And the rate at which the separation occurs is:
\begin{equation}
 \frac{d\Phi}{dt'} = \frac{2\beta}{y\sqrt{1+\frac{2\beta^2}{y}t'}} =
 \frac{2\beta}{y\left(1+\beta\phi\right)}
\end{equation}
Defining $t'_0 = -t{\rm min}$, we can write:
\begin{equation}
 \frac{d\Phi}{dt'} = \sqrt{\frac{2}{y\Delta t'}}
\end{equation}
where $\Delta t' = t' + t'_0$, the apparent age of the quasar.

As shown before in equation~(\ref{eqn.8}), the red shift $\z$ depends
on the real speed $\beta$ as:
\begin{equation}
1 \,+\, \z \,=\, \left|1 \,-\, \beta\cos\theta\right| \,=\, \left|1
  \,+\, \beta\sin\phi\right|\label{eqn.10}
\end{equation}
There are two solutions for $\z$.  For $\sin\phi < -1/\beta$,
you get
\begin{equation}
 \z_1 = -2 - \beta\sin\phi_1
\end{equation} and for $\sin\phi > -1/\beta$,
\begin{equation}
 \z_2 = \beta\sin\phi_2
\end{equation}
Thus, we get the difference in the red shift between the two hot spots
as:
\begin{equation}
 \Delta\z \approx 2 + \beta(\phi_1+\phi_2)
\end{equation}
But $\phi_1+\phi_2 = -2/\beta$ and hence $\Delta\z = 0$.  The two hot
spots will have identical red shifts, if terms of $\phi^3$ and above
are ignored.

\subsection{Time Evolution of Object Size  and Red Shift}

Fig.~\ref{PhiVsTp} shows the apparent positions ($\phi$) and the size
of the superluminal object as the observer sees it, as a function of
the observer's time ($t'$).  Fig.~\ref{zVsTp} is a similar time
evolution of the red shift ($\z$).  In this section, we describe how
these two plots are created.  It is easiest to express the quantities
parametrically as a function of the real time $t$. Referring to
Fig.~\ref{dd}, we write,
\begin{eqnarray}
  x &\,=\,& \beta t \\
t' &\,=\,& t + \sqrt{\beta^2t^2 + y^2} - y\\
\sin\phi &\,=\,& \frac{\beta t}{\sqrt{\beta^2t^2 + y^2}}
\end{eqnarray}
The solid parabola in Fig.~\ref{PhiVsTp} is $\phi$ vs. $t'$ from these
equations as $t$ is varied between $-40$ and $20$ years, with $y =
1\,000\,000$ light years and $\beta = 300$.

In order to get the variation of the size of the object (the shaded
region in Fig.~\ref{PhiVsTp}), we assume a diameter $d = 500$ light
years.
\begin{eqnarray}
t'_\pm &=& t + \sqrt{\left(\beta t \pm \frac{d}{2}\right)^2 + y^2} - y\\
\sin\phi_\pm &=& \frac{\beta t \pm \frac{d}{2}}{\sqrt{\left(\beta t \pm
    \frac{d}{2}\right)^2 + y^2}}
\end{eqnarray}
The boundaries of the shaded region are given by $\phi_+$ vs. $t'_+$
and $\phi_-$ vs. $t'_-$.

As shown before (see equation~(\ref{eqn.10}), the red shift $\z$
depends on the real speed $\beta$ as:
\begin{equation}
1 \,+\, \z \,=\,\left|1 \,+\, \beta\sin\phi\right| \,=\, \left|1 \,+\,
  \frac{\beta^2t}{\sqrt{\beta^2t^2 + y^2}}\right|
\end{equation}
Since we know $\z$ and $t'$ functions of $t$, we can plot their
inter-dependence parametrically.  This is shown in Fig.~\ref{zVsTp},
for two sets of values for $\beta$ and $y$.

\section{Apparent Expansion of the Universe}
The apparent recessional speed is the longitudinal component of
$\beta'$ is $\beta'_\parallel = \beta' \cos\theta$.  From
equation~(\ref{eqn.1}), we can see that
\begin{equation}
\lim_{\beta\to\pm\infty} \beta'_\parallel \,=\, -1
\end{equation}
The apparent recessional speed (which can be measured using red
shifts) tends to $c$ (or, $\,\beta'_\parallel\,\to\,-1$), when the
real speed is highly superluminal.  This limit is independent of the
actual direction of motion of the object $\theta$.  Thus, whether a
superluminal object is receding or approaching (or, in fact, moving in
any other direction), the appearance from our perspective would be an
object receding roughly at the speed of light.  This appearance of all
(possibly superluminal) objects receding from us at strictly
subluminal speeds is an artifact of our perception, rather than the
true nature of the universe.

\bibliographystyle{aa}
\bibliography{refs}

\end{document}